\documentclass{moriond}
\pdfoutput=1

\def\be{\begin{equation}}
\def\ee{\end{equation}}
\def\bea{\begin{eqnarray}}
\def\eea{\end{eqnarray}}

\usepackage{amsmath, amssymb}
\usepackage{xspace}

\usepackage[backend=biber, sorting=none, style=numeric-comp]{biblatex}
\usepackage{subcaption}
\graphicspath{{.}}

\newcommand\abra{ABRACADABRA\xspace}
\newcommand\cm{-10\,cm\xspace}
\newcommand{\beqg}{\be\begin{gathered}}
\newcommand{\eeqg}{\end{gathered}\ee}
\newcommand{\gayy}{g_{a\gamma\gamma}}
\newcommand{\eq}[1]{Eq.\,\ref{eq:#1}}
\newcommand{\fig}[1]{Fig.\,\ref{fig:#1}}
\newcommand{\sect}[1]{Sec.\,\ref{sec:#1}}

\newcommand{\Photo}{}

\begin{document}
\title{First Results from \abra\cm:\\A Search for Low-Mass Axion Dark Matter}

\author{ C. P. Salemi\\on behalf of the \abra collaboration}

\address{Laboratory for Nuclear Science, Massachusetts Institute of Technology, Cambridge, MA 02139, U.S.A.\vspace{2ex}}

\maketitle\abstracts{The presence of dark matter provides some of the most tangible evidence for the existence of physics beyond the Standard Model.  One compelling dark matter candidate is the axion, a light boson that was originally postulated as a solution to another outstanding issue, the strong CP problem in QCD.  \abra is an experimental program to search for sub-$\mu$eV axion and axion-like dark matter.  It searches for axion-induced modifications to Maxwell's equations with a toroidal magnet and SQUID magnetometer.  This contribution will present the first results from the prototype detector, \abra\cm.}

\section{Introduction}

The existence of dark matter is some of the strongest evidence we have for physics beyond the Standard Model.  There is strong observational evidence for matter that interacts gravitationally but that cannot be any of the visible particles from the Standard Model \cite{PDG2018}.  However, so far there have been no confirmed reports of direct detection of any such dark particles.

Theories abound as to the identity of the dark matter, and so experiments aim to look for the most highly motivated candidates.  One of the foremost such candidates is the axion, a particle originally hypothesized as a part of a solution to another outstanding problem in physics, the strong CP problem.  This is the problem that we expect QCD to violate CP symmetry, but it does not.  Instead, the CP-violating term in the QCD Lagrangian is suppressed by coefficient $\bar{\Theta}\lesssim10^{-10}$ where theoretically $\bar{\Theta}\in[0,2\pi]$, indicating a huge fine-tuning of the coefficient down to zero.

In order to remedy this, Roberto Peccei and Helen Quinn proposed a new, global $U(1)$ symmetry \cite{Peccei1977}.  When the symmetry spontaneously breaks at some large energy scale, $f_{a}$, a goldstone boson, $a$, is produced, which dynamically drives the $\bar{\Theta}$ term to zero, naturally allowing QCD to conserve CP symmetry.  This is known as the PQ mechanism.

Frank Wilcek and Steven Weinberg later pointed out that an important aspect of this mechanism is the implication the existence of a massive, pseudoscalar particle, dubbed the axion \cite{Wilczek1978,Weinberg1978}.  This new particle makes a great dark matter candidate because it is neutral and cold and can be produced in the correct abundance.  At late times, assuming it accounts for the full dark matter density, the axion field oscillates as
	\beqg\label{eq:axionOsc}
		 a(t)=\frac{\sqrt{\rho_{DM}}}{m_{a}}\sin(m_{a}t)
	\eeqg
where $\rho_{DM}$ is the dark matter energy density and $m_{a}\propto1/f_{a}$ is the mass of the axion.  This oscillating field of axions leaves us a signature to detect.

\section{Axion Detection with \abra}

\subsection{Axion Electromagnetism}

When writing down the axion-modified Standard Model Lagrangian, we must include all of the possible interactions of the axion with other particles.  The number and strength of such possible interactions depends on model, and in particular there are two leading models, KSVZ and DFSZ \cite{Kim1979,Shifman1980,Dine1981,Zhitnitsky1980}.  However, in all possible models, the couplings are proportional to $1/f_{a}$ and so are suppressed by the large energy scale of PQ symmetry breaking.

Most experiments, including \abra, use axions' coupling to photons to search for them.  Axions modify QED as
	\be \mathcal{L}\supset\frac{1}{4}g_{a\gamma\gamma}aF_{\mu\nu}\tilde{F}^{\mu\nu}=-g_{a\gamma\gamma}a{\bf E}\cdot{\bf B} \ee
where $\gayy$ is the coupling strength of the axion to two photons.  We can use this to modify the sourceless Amp\`ere's law to the form
	\beqg
		\nabla\times{\bf B}=\frac{\partial{\bf E}}{\partial t}-g_{a\gamma\gamma}\left({\bf E}\times\nabla a-\frac{\partial a}{\partial t}{\bf B}\right)
	\eeqg
We take it in the limit that $\nabla a\to0$ because we expect the axion field to be quite uniform over the distance scale that our detector probes.  The rightmost, axion-induced term has the same time-dependent form of the classical effective current term and so we can rewrite the axion interaction as an effective current,
	\beqg\label{eq:Jeff}
		{\bf J}_{eff}=\gayy\frac{\partial a}{\partial t}{\bf B}=\gayy\sqrt{2\rho_{DM}}\cos(m_{a}t){\bf B}
	\eeqg
where the second step comes from plugging in the axion field from \eq{axionOsc}.

\subsection{\abra Detection Concept}\label{sec:abraConcept}

\begin{figure}[h]
	\centering
	\begin{subfigure}[b]{0.5\textwidth}\centering
		\includegraphics[width=0.6\textwidth]{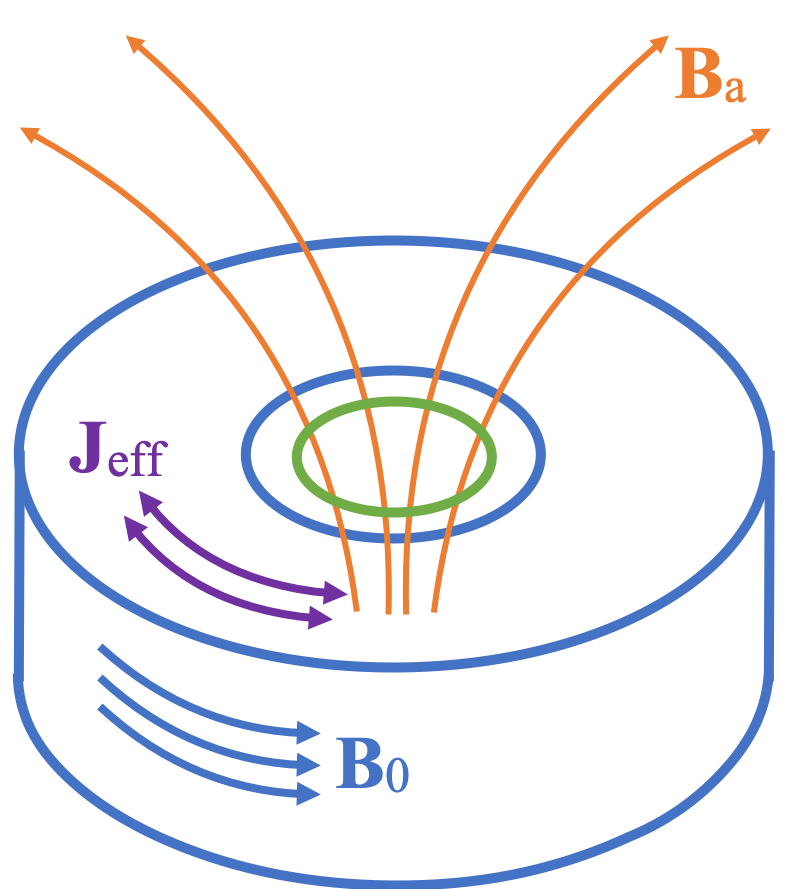}
	\caption{\label{fig:detDiagram}}
	\end{subfigure}\hfill%
	\begin{subfigure}[b]{0.5\textwidth}\centering
		\includegraphics[height=0.1\textheight]{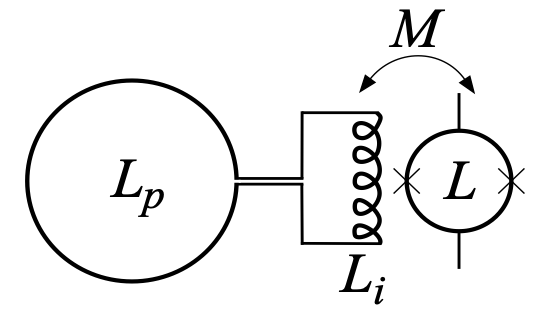} \\
		\includegraphics[height=0.1\textheight]{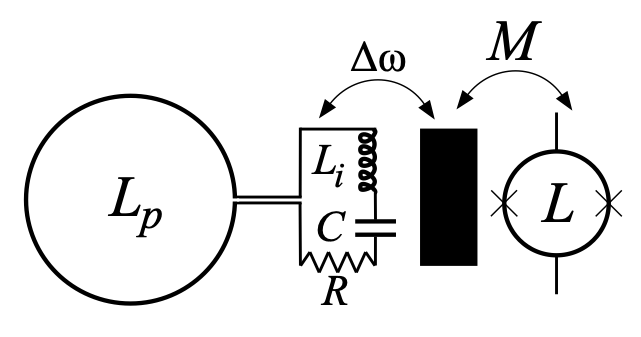}
	\caption{\label{fig:readout}}
	\end{subfigure}
	\caption{$(a)$, diagram of the \abra detector.  The toroid's magnetic field ($\bf B_{0}$, blue) sources an oscillating effective axion current ($\bf J_{eff}$, purple).  This effective current then creates a real oscillating magnetic field ($\bf B_{a}$, orange) in the center of the toroid.  These changes in magnetic flux are detected with a pickup circuit (green).  $(b)$, the pickup circuit coupled to a SQUID amplifier in, top, broadband mode and, bottom, resonant mode.  Taken from \cite{Kahn2016}.}
\end{figure}

\abra (A Broadband/Resonant Approach to Cosmic Axion Detection with an Amplifying B-field Ring Apparatus) looks for the axion effective current with a toroidal magnet and superconducting sensors \cite{Kahn2016}.  The toroid's strong azimuthal magnetic field sources an effective current parallel to it given by \eq{Jeff}.  Using classical Maxwell's equations, we can see (see \fig{detDiagram}) that this oscillating azimuthal effective current sources an oscillating real magnetic field in the central region of the toroid where there would otherwise be no field.  This change in the magnetic flux through the center can be detected with a pickup circuit.  The pickup must be superconducting with highly sensitive amplifiers because the expected signal is very small due to the feeble coupling ($\gayy$) of the axion field to the supplied magnetic field.

Signal readout can be done in two modes: broadband and resonant (see \fig{readout}).  The first involves reading out all frequencies simultaneously.  In this scenario, the only amplification is that done by the sensor-amplifier (a SQUID in \abra\cm) that is directly coupled to the pickup.  The second mode has the addition of an LC resonator in the pickup circuit.  Signals at the resonance frequency are additionally enhanced by the Q-value of the resonator, with Q's of up to $\sim10^{6}$ being reasonable.  However, this requires varying the resonator frequency to scan over the possible axion masses.  So, with the limiting factor being time, there is a trade-off between averaging for a long time over a wide range of frequencies (broadband) and scanning slowly over many axion masses (resonant).

With both readout methods the resulting signal is time series data that we Fourier transform to look for peaks in frequency space corresponding to the axion mass.  The axion signal peak is very narrow, $\Delta f/f\approx10^{-6}$, because of the coherence of the axion field.  The shape of the signal is determined by the velocity distribution of the axions; with the Standard Halo Model we expect a Maxwell-Boltzmann distribution.

\section{\abra\cm}

\subsection{Detector}

The prototype detector, \abra\cm, is a proof-of-concept located at MIT to demonstrate the efficacy of the method detailed in \sect{abraConcept}.  The magnet toroid is 12\,cm in diameter by 12\,cm in height and is mounted in a tin spin-coated copper can that is hung on the bottom plate of an Oxford Instruments Triton 400 dilution refrigerator.  The magnet is cooled to 1.2\,K.  In the center of the magnet a 2.0\,cm radius loop of 1\,mm diameter solid NbTi wire is mounted as a pickup loop.  Its leads are run out of the shield can and up to the SQUIDs, which are on a higher plate and kept at 870\,mK.  We calibrate the detector via another wire that lies in the magnetic field of the toroid to mimic an axion signal.  We feed signals into it via 90\,dB of attenuation in order to simulate the tiny axion signal, as well as to isolate the calibration circuit from environmental noise.

\abra\cm's first physics run was last summer 2018.  We calibrated the detector both before and after the run and at a range of frequencies and amplitudes with the magnet on and off.  The calibrations were consistent with each other, but we found that the detector sensitivity was a factor of $\sim6.5$ lower than expected.  We believe that this is primarily due to parasitic inductance in the twisted pair wire that connects the pickup loop to the SQUIDs.  This will be fixed in the next run of \abra\cm, as explained in \sect{nextSteps}.

\subsection{Data Collection and Analysis}

Over the course of the four week run, the DAQ sampled continuously at 10\,MS/s for a total of 25 trillion samples.  The time series data was Fourier transformed and averaged in $T_{int}=10$\,s buffers, creating a PSD that we call $\bar{\mathcal{F}}_{10\rm M}$.  The time series data was also downsampled to 1\,MS/s, transformed, and averaged in $T_{int}=100$\,s buffers to create a second PSD, $\bar{\mathcal{F}}_{1\rm M}$.  The frequency range of $\bar{\mathcal{F}}_{10\rm M}$ ($\bar{\mathcal{F}}_{1\rm M}$) is between $f=1/T_{int}=100$\,mHz ($10$\,mHz) and the Nyquist frequency, $f=5$\,MHz (500\,kHz).  The frequency resolution of each data set is also given by $\Delta f=1/T_{int}$.  Because of their different frequency ranges and resolutions, the two PSDs are used for the analysis of different regions of axion mass.

\begin{figure}[h]
	\begin{center}
	\includegraphics[width=\textwidth]{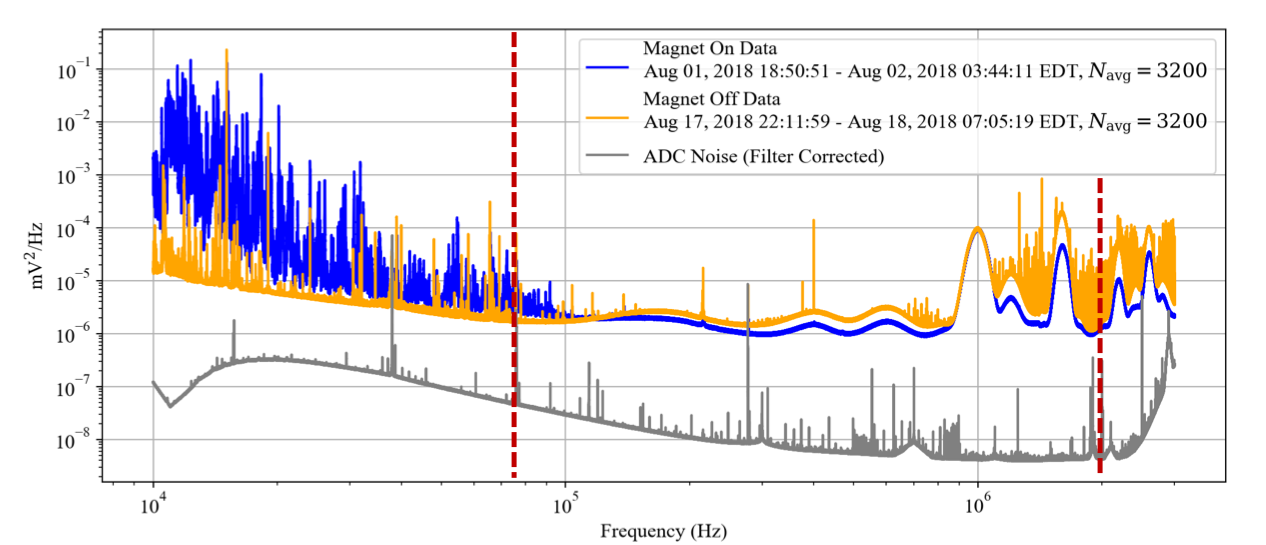}
	\caption{Averaged flux spectrum from the magnet-on run and magnet-off run, plotted in blue and orange respectively, with an averaged spectrum from a digitizer-only run in grey.  The axion search region is between the dashed red vertical lines.  Reproduced from \cite{Ouellet2019b}.}
	\label{fig:spectrum}
	\end{center}
\end{figure}

An averaged flux spectrum from the run is shown in \fig{spectrum}, plotted with averaged spectra from a magnet-off run and a digitizer-only run.  On the low end of the flux spectra there is elevated noise that increases in magnitude with decreasing frequency.  This we can attribute to stray fields from the magnet and the environment that vibrate through our pickup loop.  As can be seen by comparing the blue and orange curves, these vibration-induced lines significantly decrease when the magnet is off.  To check that the low frequency noise was indeed due to vibration, we mounted an accelerometer on the top of the fridge.  We saw a strong correlation between the accelerometer and pickup flux peaks in the region where our accelerometer was sensitive, below 10\,kHz, and believe that the forest of lines below around 100\,kHz is also likely due to vibration.  In order to fit the data into our digitizer window, we needed to filter away some of the noise at low frequencies and so added a 10\,kHz high-pass filter.

On the high frequency end of the spectrum are a series of wide bumps.  We observe these in both magnet-off and magnet-on runs, as well as when we take data from the SQUIDs with no pickup loop attached to their input.  They are likely due to either a ground loop or environmental noise that is entering the system between the SQUIDs and the DAQ.  These bumps are wide enough that although they reduce our sensitivity there, we can fit our background on top of them.

In the middle region between $\sim75\,$kHz$-1\,$MHz, there is a wide flat region with few peaks.  The noise here sits at $\sim1\,\mu\Phi_{0}$, our expected SQUID flux noise floor.  The slight difference in magnitude between the magnet-on and magnet-off spectra corresponds to an expected drift over time of total SQUID flux noise.

The region between the dashed red vertical lines is where we performed an axion search, between $75\,$kHz$-2\,$MHz or equivalently $0.31-8.3\,$neV.  The lower bound was fixed to remove most of the vibrational noise and the upper bound was set by the roll-off frequency of a 1.9\,MHz low-pass filter that we added for anti-aliasing.

In addition to frequency cuts, we made time cuts because certain frequencies exhibited transient noise.  There were two periods when this took place, at the beginning of the run and then in the middle of the run for a couple of days after we went into the lab to do a nitrogen refill.  The noise during these periods had many sharp peaks which could mimic an axion signal.  However, the axion signal should be constant over the entire run.  To reduce the signal contamination by these noisy periods, we removed spectra from the total dataset by placing cuts on the total number of $3\sigma$ signals in each spectrum.  Only spectra with less than thirty $3\sigma$ detections were included in the axion search, which resulted in an $\sim70\%$ efficiency.

The final set of cuts that we implemented allowed us to veto persistent environmental peaks.  This type of background is also present when the magnet is off, which indicates that it cannot be from an axion.  For this, the magnet-off data was processed using the same method as that with the magnet on.  Any signals that passed the $5\sigma$ threshold in the magnet-off dataset were vetoed in the magnet-on dataset.

For each mass point in our PSDs, we calculate a likelihood function, $\mathcal{L}$, in order to perform a log-likelihood ratio test between the best fit and the null hypothesis.  This was done as a joint fit over the 53 (24) averaged spectra in the $\bar{\mathcal{F}}_{10\rm M}$ ($\bar{\mathcal{F}}_{1\rm M}$) data sets.  For our discovery search we set a $5\sigma$ threshold on our test statistic,
	\beqg
		{\rm TS}(m_{a})=2\ln\left[\frac{\mathcal{L}\left( {\bf d}_{m_{a}}|\hat{A},{\bf\hat{b}} \right)}{\mathcal{L}\left( {\bf d}_{m_{a}}|A=0,{\bf\hat{b}}_{A=0}\right)}\right]
	\eeqg
where $\bf d$ is our data, $A$ is the signal strength, $\bf b$ is the background, and the hatted parameters are those that maximize the likelihood.  Note that our threshold accounts for the Look Elsewhere Effect because the spectra include $\sim8.1\times10^{6}$ mass points.  Where we did not observe a signal we set 95\% confidence level limits with a similar test statistic.

\subsection{Results}

\begin{figure}[h]
	\begin{center}
	\includegraphics[width=\textwidth]{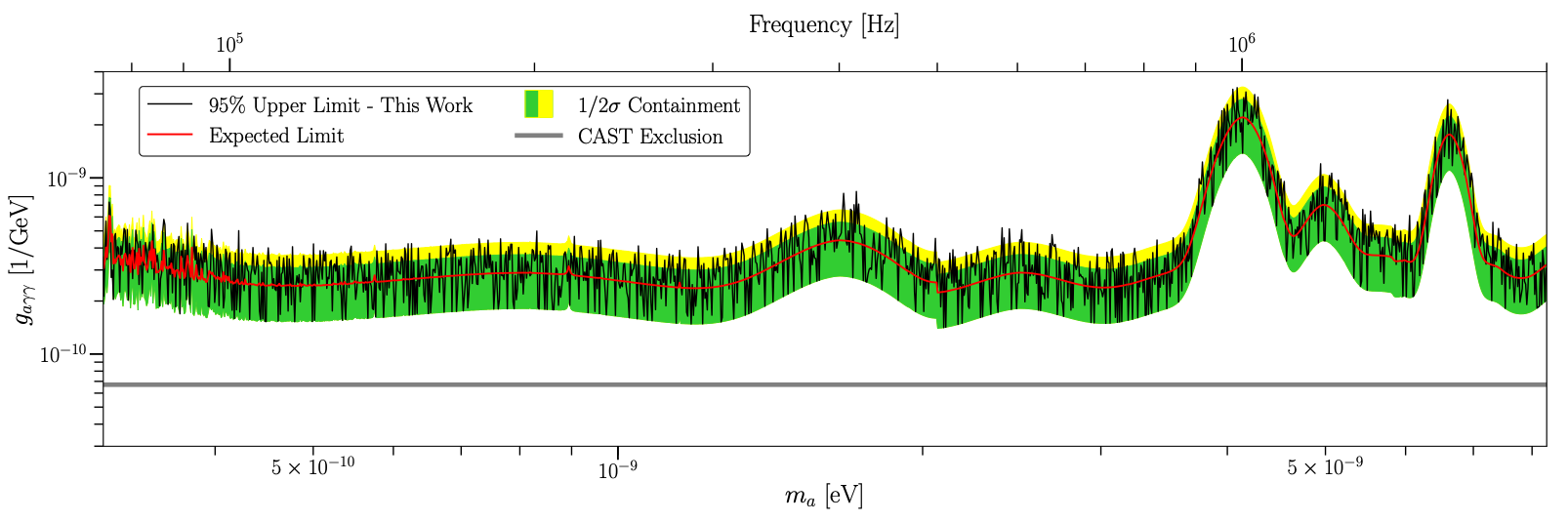}
	\caption{95\% C.L. limit in black with the 1 and 2$\sigma$ containment in green and yellow respectively.  The red curve shows the expected limit and the grey marks the limits on solar axions set by CAST.  Reproduced from \cite{Ouellet2019a}.}
	\label{fig:limits}
	\end{center}
\end{figure}

We found no $5\sigma$ excesses in the data from this first, month-long run.  We set 95\% C.L. limits on axion dark matter as shown in \fig{limits} that are competitive with the limits on solar axions set by CAST \cite{Anastassopoulos2017}.

\section{Next Steps}\label{sec:nextSteps}

The next step for \abra\cm is another month-long run with modifications intended to improve or eliminate the factor of 6.5 reduction in sensitivity.  To this end, we are currently implementing changes in the readout circuit, modifying the geometry of the pickup loop and shortening the twisted pair wires that run from the pickup to the SQUIDs.  This run should allow us to surpass the limits set by CAST.

After \abra\cm's second run, the detector will be repurposed as an R\&D testbench for future generations of \abra.  The next generations will be larger detectors with higher magnetic fields, and they will primarily use the resonant readout method.  The next generation will be a 40\,cm diameter, 5\,T toroidal magnet that should be able to probe into the QCD axion regime for axions with masses just below a $\mu$eV.  Later, meter-scale experiments will be able to expand the range down to tens of neV for QCD axions, and both 40\,cm and 1\,m magnets running in broadband mode will be able to probe a large swath of axion-like particle parameter space for sub-$\mu$eV masses.

The \abra program aims to find or exclude axions and axion-like particles over a wide range of couplings and masses below a $\mu$eV.  \abra\cm has demonstrated the promise of this method, and with a month-long run has set competitive limits on axion-like particles with masses $0.31-8.3\,$neV.  \abra\cm is continuing its search, and we are presently preparing for another physics run with improved sensitivity.
 
\section*{Acknowledgments}

This material is based upon work supported by the National Science Foundation under grant numbers NSF-PHY-1806440 and NSF-PHY-1658693 and by the National Science Foundation Graduate Research Fellowship under Grant No. 1122374.

\printbibliography[title={References}]

\end{document}